\documentclass[preprint,12pt]{elsarticle}
\usepackage{epsfig,graphics,graphicx,amsmath,amssymb,float}

\usepackage{amssymb}

\journal{Nuclear Physics B}

\begin{document}

\begin{frontmatter}



\title{ Kink properties in Lorentz-violating scalar field theory }


\author{Reza Moazzemi}\ead{r.moazzemi@qom.ac.ir}
\author{Mohammad Mehdi Ettefaghi}\ead{mettefaghi@qom.ac.ir}
\author{Amirhosein Mojavezi}\ead{amojavezi98@gmail.com}
\address{Department of Physics, University of Qom, Ghadir Blvd., Qom 371614-6611, I.R. Iran} 

\baselineskip=.75cm

\begin{abstract}
	We consider topological defects for the $\lambda\phi^4$ theory in (1+1) dimensions with a Lorentz-violating background. 
	It has been shown, by M. Barreto et al. (2006) \cite{barreto2006defect}, one cannot have original effects in (the leading order of) single scalar field model.
	Here, we introduce a new Lorentz-violating term, next to leading order  which cannot be absorbed by any redefinition of the scalar field or coordinates. 
	Our term is the lowest order term which leads to concrete  effects on the kink properties. We calculate the corrections to the kink shape and the corresponding  mass.  Quantization of the kink is performed and the revised modes are obtained. We find the bound and continuum states are affected due to this Lorentz symmetry violation.
\end{abstract}


\end{frontmatter}


 \baselineskip=.75cm
\section{Introduction}
Nonlinear field models are very interesting in various areas of physics including particle physics, condensed matter physics, cosmology and optics \cite{manton2004topological,belova1997solitons}. 
Solitons and solitary waves that have localized energy and propagate without losing the shape are important solutions of  some nonlinear field equations. We can cite the following examples of solitons; vertices in superconductors \cite{abrikosov2004nobel} and fluids \cite{kleckner2013creation}, cosmic strings \cite{vilenkin2000cosmic}, monopoles \cite{t1974magnetic} and instantons \cite{schaefer2002instanton}.
 There are also many works on supersymetric solitons, see for example \cite{shifman,opef,schon,graham}. The  kinks which appear in (1+1)-dimensional models with a potential possessing two or more degenerated minima, are the simplest examples of solitons \cite{vachaspati2006kinks}. Some realistic models in (3+1) or (2+1) dimensions can be reduced to the effective (1+1)-dimensional dynamics.  In some theories such as $\lambda\phi^4$, the existence of kink in higher dimensions is banned \cite{cuevas2019discrete}. Actually, to construct a modified kink in such a case we should interpolate from field theory in the Minkowski space-time to field theory in curved space-time background \cite{waterhouse2019phi}.

One of the most important symmetries which should be considered in theories for high energy 
physics is the Lorentz symmetry. So far, it has not been observed any deviation of this symmetry by experimental inspections. However, Lorentz symmetry may be violated due to the quantum gravity in the Plank scale \cite{kostelecky1995cpt}.  Noncommutative theories also constitute an essential motivation for study of Lorentz-breaking theories and involve a Lorentz-breaking tensor $\Theta^{\mu\nu}$ \cite{carroll2001noncommutative,aghababaei2017lorentz}.
Therefore, it is interesting to restudy a Lorentz covariant theory when a small violation in Lorentz symmetry is allowed. The standard model and QED, for instance, were extended by the LV terms \cite{colladay1998lorentz}. The Casimir effect in an LV theory is also considered in some works, see for example \cite{santos,cruz,moazzemi}.

 So far, various issues and problems have been studied within the LV extended standard model framework (for instance see \cite{tasson} and references therein). This investigations have been done with basically a
twofold purpose: the determination of new physical effects induced by the Lorentz-violating coefficients and the imposition of stringent upper bounds on the LV coefficients that constrain the magnitude of Lorentz breaking. Our goal in this paper is to study the former. Such studies have been done in several papers such as Refs. \cite{bailey,ferr1,ferr2}. In Ref. \cite{bailey}, authors have investigated the static limit of Lorentz-violating electrodynamics in vacuum and in media and  obtained explicit solutions for some simple cases. More details of this theory have been investigated within the Refs. \cite{bailey,ferr1} and classical solutions are also properly examined \cite{ferr3,ferr4}.

In addition, LV effects on topological defects are considered e.g. in Refs. \cite{gopakumar2000noncommutative,belich2005non,seifert2013lorentz,lubo2005global} and, in particular, for scalar field theories in Refs. \cite{passos2018soliton,bazeia2010lorentz,barreto2006defect}. In Ref. \cite{barreto2006defect},  authors considered two models in (1+1) dimensions (with one and two scalar fields) in which the Lorentz and CPT symmetries are violated. Actually, in the case of a single scalar field, the LV term, modeled by $-\frac{1}{2}k^{\mu\nu}\partial_\mu\phi\partial_\nu\phi$, where $k^{\mu\nu}$ is LV parameter, can be removed through a linear transformation of coordinates, such a transformation will generate a Jacobian multiplier in the new action. In fact, one can imagine $k^{\mu\nu}=cu^{\mu}u^{\nu}$ characterized by the LV vector $u^{\mu}$ which is a spacelike two-vector denoting the  aether \cite{carrol2008,gomes}. Therefore, in (1+1) dimensions, we can say that there does not exist any observable effect due to the leading order of this CPT even Lorentz violating interaction of scalar field to the aether. Therefore, the nontrivial contributions due to this term appear when we couple the model with more sophisticated fields. However, the effects of higher order LV terms have not been yet considered. One may naturally expect that there must exist some effects due to more complicated LV interacting terms. In this paper, we investigate the unremovable LV effects that occurred in the next to leading order.  Explicitly, we study the  LV interacting scalar field theory modeled by $-\frac{1}{2}c\phi^2 (u\cdot\partial\phi)^2$  in (1+1) dimensions. This is the lowest order term which leads to concrete  effects on the kink properties. We discuss two different forms of LV terms $cu^\mu$, time-like and space-like, i.e. Lorentz symmetry breaking in the direction of time and space, respectively. For the case of the time-like vector, we see there is no concrete effect. Therefore, we study only the space-like LV vector.

After a very brief review of the standard kink in Sec. \ref{sec2}, we first, find the modification to the shape of the kink in subsection \ref{2b} and then calculate the mass correction due to this LV term. The stability equation is obtained in Sec. {3a} and based on this new corrected stability equation we find the modification of bound and continuum states. The last section is devoted to summarizing the results and conclusions.

\section{Classical kinks in Lorentz violating theory}\label{sec2}

Here we first, very briefly, explain the standard kink and derive its mass. The appropriate Lagrangian for spontaneously broken symmetry in 1+1 dimensions within the real scalar $\phi^4$ theory is:
\begin{equation}\label{a1}
	{\cal L}(x)
	=\frac{1}{2}(\partial_{t}\phi)^{2}-\frac{1}{2}(\partial_{x}\phi)^{2} -\frac{\lambda}{4}\left(\phi^2- \frac{m^2}{\lambda}\right)^2.
\end{equation}
By making the following scaling:
\begin{equation}
	\phi\rightarrow\frac{\sqrt{m}}{\lambda}\phi,\hspace{1cm}
	x\rightarrow \frac{x}{m},\hspace{1cm} 
	{\cal L}\rightarrow \frac{m^4}{\lambda}{\cal L},
\end{equation}
the dimensionless Lagrangian is achieved as follows:
\begin{equation}\label{a}
	{\cal L}(x)
	=\frac{1}{2}(\partial_{t}\phi)^{2}-\frac{1}{2}(\partial_{x}\phi)^{2} -\frac{1}{4}(\phi^2- 1)^2.
\end{equation}
Equation of motion for Lagrangian \eqref{a} reads as
\begin{equation}\label{b}
	\partial^2_{t}\phi-\partial^2_{x}\phi-\phi=-\phi^3.
\end{equation}
This equation has four static solutions: two trivial non-topological solutions $\phi_{\rm{tri.}}(x)=\pm1$, and two topological ones (kink and anti-kink):
\begin{equation}\label{kinkcl}
	\phi_{\rm{ k}}(x)=\pm\tanh\left(\frac{x-x_0}{\sqrt{2}}\right),
\end{equation}
which approach two different values as $x\to\pm\infty$ corresponding to the two possible vacuum states.
(For more details see please \cite{Rajaraman1982}). \\
Putting the solutions \eqref{kinkcl} in the related Hamiltonian  
\begin{equation}
	H=\int\left[\frac{1}{2}(\partial_t\phi)^2+\frac{1}{2}(\partial_x\phi)^2+\frac{1}{4}(\phi^2- 1)^2\right]dx,
\end{equation}
one can find the familiar  classical mass of the kink as
\begin{equation}
	M_{\rm{cl.}}=\frac{2\sqrt{2}}{3}.\label{classic mass}
\end{equation}

\subsection{$\phi^4$ theory with Lorentz symmetry violation  }\label{2b}

In order to have an LV theory (with preserving the other favorite symmetries such as parity and $Z_2$), one may add the simplest possible term  $-\frac{1}{2}k^{\mu\nu}\partial_\mu\phi\partial_\nu\phi$ to the standard Lagrangian \eqref{a}. However, as we stated in the Introduction, this term  gives no real manifestation of Lorentz violation. Note that, since the existence of the kink owes the interacting term in the Lagrangian, we expect an LV term in the interaction part might affect the kink directly. Therefore, we introduce a model in which a bit more complicated LV term is added to the Lagrangian, as follows:
\begin{equation}\label{c}
	{\cal L}(x)
	=\frac{1}{2}(\partial_{t}\phi)^ {2}-\frac{1}{2}(\partial_{x}\phi)^{2} - \frac{1}{2}c\phi^2 (u\cdot\partial\phi)^2 -\frac{1}{4}(\phi^2- 1)^2,
\end{equation}
where the vector $cu^\mu$ manifests a preferred direction in the space-time which is responsible for Lorentz symmetry violation.
Here, we consider two different types of  Lorentz symmetry violation; first, a violation occurs in the direction of time, called time-like (TL), and second in the direction of space, called space-like (SL).

\subsubsection{TL vector case}
In the case of the TL vector, $u^\mu=(1,0)$, the third term in \eqref{c} is $\frac{1}{2}c\phi^2 (\partial_t\phi)^2$. Hence, the equation of motion \eqref{b} is modified as:
\begin{equation}
	\partial^2_{t}\phi-\partial^2_{x}\phi-c\phi(\partial_t\phi)^2-c\phi^2\partial^2_{t}\phi-\phi=-\phi^3.
\end{equation}
It is obvious that the static solutions of this differential equation are not different with \eqref{b}. Accordingly, Lorentz violation in the time direction has no effect on the kink properties.

\subsubsection{SL vector case}
\begin{figure}
	\begin{center} \includegraphics[width=13cm]{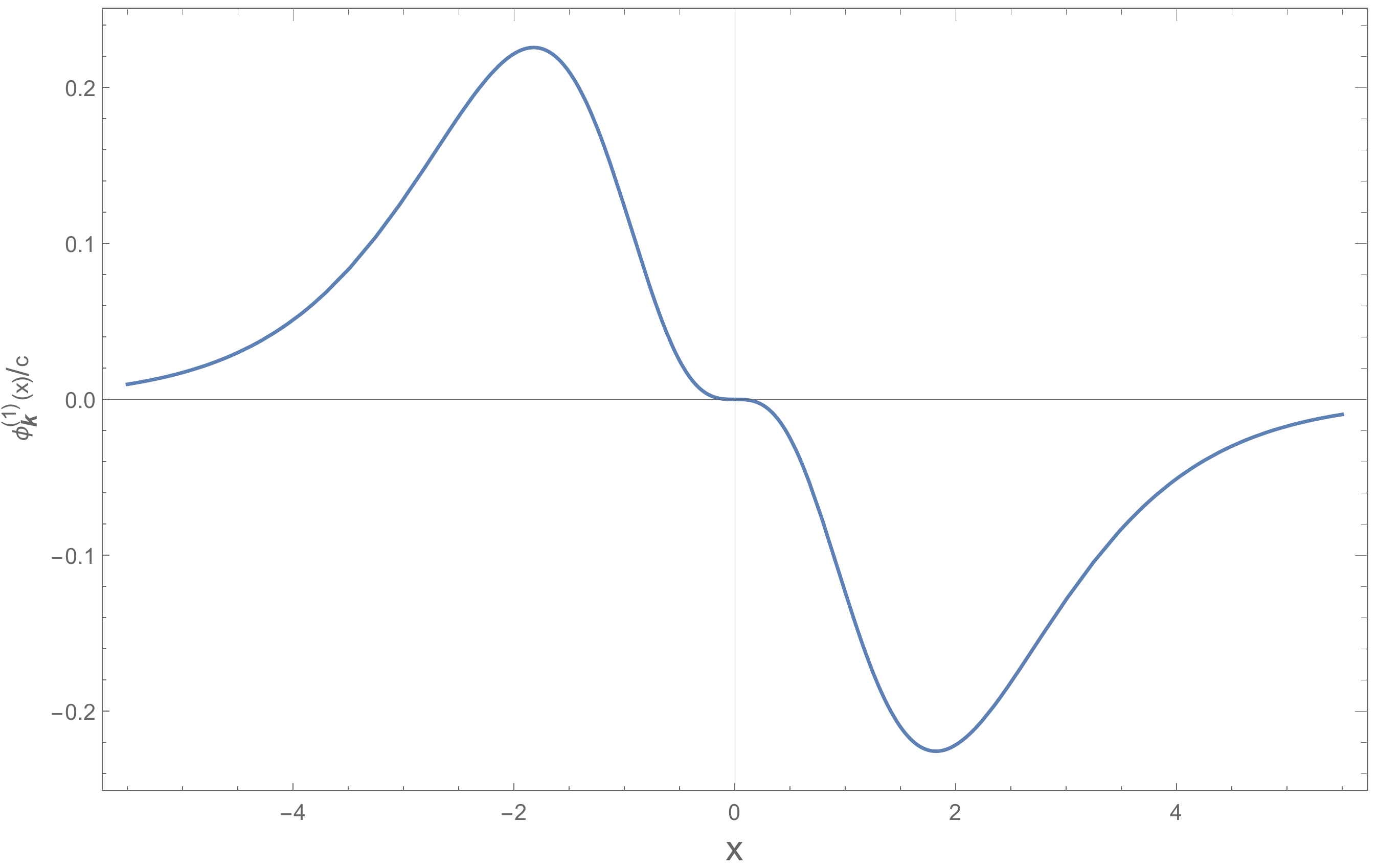} \caption{ {\small First order correction to the kink function due to Lorentz violation} \label{fig1}}
	\end{center}
\end{figure}

In the SL case, $u^\mu=(0, 1)$ the LV term in \eqref{c} becomes $-\frac{1}{2}c\phi^2 (\partial_x\phi)^2$ and equation \eqref{b} changes to
\begin{equation}\label{em2}
	\partial^2_{t}\phi-\partial^2_{x}\phi-c\phi(\partial_x\phi)^2-c\phi^2\partial^2_{x}\phi-\phi=-\phi^3.
\end{equation}
One can solve this differential equation using the perturbation method, i.e.  substituting $\phi_{\rm{k}}(x)=\phi^{(0)}_{\rm{k}}(x)+c\, \phi^{(1)}_{\rm{k}}(x)+c^2\phi^{(2)}_{\rm{k}}(x)+\dots$ in Eq. \eqref{em2}, the solutions are derived as 

\begin{eqnarray}
	\nonumber\phi_{\rm{k}}(x)&=&\phi^{(0)}_{\rm{k}}(x)+c\, \phi^{(1)}_{\rm{k}}(x)\\&=&\pm\bigg\{\tanh\left(\frac{x-x_0}{\sqrt{2}}\right)\nonumber \\&&\hspace{1cm}- c\ \frac{1}{2}  {\rm{sech}}\left(\frac{x-x_0}{\sqrt{2}}\right)\left[\frac{x-x_0}{\sqrt{2}}- \tanh\left(\frac{x-x_0}{\sqrt{2}}\right)\right] \bigg\},\label{ki}
\end{eqnarray}
where we retain terms up to order $c$. In Fig. \ref{fig1}  the correction term due to the Lorentz violation is plotted. The related Hamiltonian for this case is
\begin{equation}
	H=\int\left[\frac{1}{2}(\partial_t\phi)^2+\frac{1}{2}(\partial_x\phi)^2+\frac{1}{2}c\phi^2 (\partial_x\phi)^2+\frac{1}{4}(\phi^2- 1)^2\right]dx.
\end{equation}
Inserting (\ref{ki}) in this Hamiltonian, we obtain
\begin{equation}
	M_{\rm{cl.}}=\frac{2\sqrt{2}}{3}+ \frac{\sqrt{2}}{15} c+{{\cal{O}}(c^2)}.
\end{equation}
This is the classical mass of the kink in the LV theory up to the first order of Lorentz violation parameter $c$ . 

\section{Quantizing the Kinks}\label{3a}

As we know, the trivial vacuum and kink are classical static solutions of a nonlinear relativistic field equation. In principle, they can be associated with quantum extended particle states. Therefore, we are now going to study the relevance of them to the corresponding quantum field theory. To do this, it may be thought that one needs either the Feynman path-integral or the canonical quantization methods. However, another simple method which is used in the literature is an approach through which one expands the energy corresponding to the static states via a semiclassical method. Then, using the classical solutions one can obtain a little information about the quantum behavior of the system. This semiclassical
quantization technique has been firstly introduced for relativistic field theories, using an appropriate generalization of the WKB approximation in quantum mechanics by Dashen,
Hasslacher and Neveu \cite{Dashen1974,Dashen1974a}. For future use, here, we very briefly review this approach and some of its conclusions.

The energy of static states involves only the potential energy which can be expanded about a minimum as follows: 
\begin{equation}\label{11}
	V[\phi]=V[\phi_0]+\int dx\frac 1 2 \left[\zeta(x)\left(-\nabla^2+\left.\frac{d^2{\cal U}}{d\phi^2}\right|_{\phi_0(x)}\right)\zeta(x)\right]+\dots,
\end{equation}
where $\zeta(x)=\phi(x)-\phi_0(x)$ and $\cal U$ is the interacting potential density along with the mass term. In fact, this is  functional Taylor expansion of $V[\phi(x)]$ about the static solution $\phi_0(x)$. In Eq. \eqref{11}, integration by part has been used and dots represent cubic and higher terms which can be treated by perturbation. This method can be used when the weak coupling approximation is established. The first term in Eq. \eqref{11} is related to classical contribution in energy.  The second term is the leading order quantum correction which is similar to the harmonic oscillator with  well-known quantum physics. We can find the normal modes by solving the following stability equation:
\begin{equation}\label{eqmo}
	\left(-\nabla^2+\left.\frac{d^2{\cal U}}{d\phi^2}\right|_{\phi_0(x)}\right)\zeta_n(x)=\omega^2_n\zeta_n(x).
\end{equation}
Now, one can construct the energy spectrum of a static state corrected by quantum fluctuations as follows:
\begin{equation}
	E_{\{n_i\}}=V[\phi_0]+\sum_i\left(n_i+\frac 1 2 \right)\omega_i,
\end{equation}
where $n_i$ is the excitation number of the $i$th normal mode. Using the Lagrangian  Eq. \eqref{a},  the Eq. \eqref{eqmo} reduces to 
\begin{equation}\label{fff..}
	\bigg[-\partial^2_{x}-1+3\phi_{0}^2(x) \bigg]\zeta_n(x)=\omega_n^2\zeta_n(x).
\end{equation}
For the trivial vacuum $\phi_0=\phi_{\rm{tri.}}=\pm1$, this equation has plane wave solutions $e^{\pm ikx}$ on which we can impose the Periodic Boundary Condition (PBC) in a box with length $L$. Therefore, $\omega_n^2=k_n^2+2$ with $k_n=2n\pi/L$, and the energy becomes
\begin{equation}\label{vac..}
	E_{\{n_i\}}^{\rm{trv.}}=\sum_{i=-\infty}^{+\infty}\left(n_i+\frac{1}{2}\right)\sqrt{k_i^2+2}.
\end{equation}

For the $\phi_0=\phi_{\rm{k}}=\pm\tanh\left(\frac{x-x_0}{\sqrt{2}}\right)$, the Eq. \eqref{fff..}  becomes 
\begin{equation}\label{fff...}
	\bigg[-\partial^2_{x}-1+3\tanh^2\left(\frac{x}{\sqrt{2}}\right) \bigg]\zeta_n(x)=\omega_n^2\zeta_n(x),
\end{equation}
where we choose $x_0=0$. The Eq. \eqref{fff...} has two bound state solutions  (regardless of normalization factors)  
\begin{equation}\label{BSs}
	\left\{\begin{array}{cc}
		\zeta_0(x)=\frac{1}{\cosh^2(x/\sqrt{2})}&\mbox{with}\quad\omega_0=0,\\
		\zeta_1(x)=\frac{\sinh(x/\sqrt{2})}{\cosh^2(x/\sqrt{2})}&\qquad\mbox{with}\quad\omega_1=\sqrt{2/3},
	\end{array}
	\right.	
\end{equation}
followed by a continuum
\begin{equation}\label{CS}
	\begin{array}{cc}& 
				\zeta_q(x)=e^{iqx/\sqrt2}\left[3\tanh^2(x/\sqrt2)-1-q^2-3iq\tanh(x/\sqrt2)\right]\\&\hspace{-1cm}		\quad\mbox{with}\qquad\omega_q=\sqrt{2+q^2/2}.
			\end{array}
\end{equation}
The zero mode is corresponding to the  translational invariance of the potential. The asymptotic behavior  of $\zeta_q(x)$ at $\pm\infty$ is $\exp[iqx/\sqrt2\pm\delta(q)/2]$ where $ \delta(q)=-2\tan^{-1}\left(\frac{3q}{2-q^2}\right)$ is the phase shift of scattering states. Again, the PBC gives the allowed mumenta 
\begin{equation}\label{pbc}
	q_n L/\sqrt2+\delta(q_n)=2n\pi,
\end{equation}
and the energy spectrum is
\begin{equation}
	E_{\{n_1,n_{q_i}\}}=\frac{2\sqrt2}{3}+\left(n_1+\frac{1}{2}\right)\sqrt{\frac{3}2}+\sum_{q_i}\left(n_{q_i}+\frac{1}{2}\right)\sqrt{2+\frac{q^2_i}2}\ .
\end{equation}
For more details please see \cite{Rajaraman1982}, for instance.

\subsection{Excitations of the trivial vacuum}
In this subsection we consider the quantization of the trivial solution in the LV theory. For this purpose,  using Lagrangian \eqref{c} we expand the potential about one of its trivial vacuum.  Similar to Eq. \eqref{11}, the functional expansion up to the second order of $\zeta$ is written as follows:
\begin{equation}
	V[\phi]=V[\phi_{\rm{tri.}}]+\int d x\left[ \frac{1}{2}\zeta(x)\left(-(1+c)\partial^2_{x}+2\right)\zeta(x)\right]+\dots.
\end{equation}
Accordingly, the eigenvalues in lowest-order quadratic terms are obtained from this equation
\begin{equation}\label{vac0} \left[-(1+c)\partial^2_{x}+2\right]\zeta_n(x)=\omega_n^2\zeta_n(x).
\end{equation}
This equation has plane wave solutions $e^{\pm ikx}$ with $\omega_n^2=(1+c)k_n^2+2$. After imposing the PBC,  we can write the energy as
\begin{equation}\label{vac}
	E_{\{n_i\}}=\sum_i \left(n_i +\frac{1}{2}\right)\omega_i=\sum_{i=-\infty}^{+\infty}\left(n_i+\frac{1}{2}\right)\sqrt{(1+c)k_i^2+2},
\end{equation}
where the classical energy $V[\phi_{\rm{tri.}}]$ is zero. We see from Eq. \eqref{vac} (or Eq. \eqref{vac0}) that  the Lorentz violation effect can be eliminated by the redefinition of $k_n$ (or spatial coordinate). Therefore, the vacuum solution does not receive any genuine correction due to the Lorentz violation in this case.

\subsection{Excitations of kink}

Now, we discuss the quantization of the kink. Similar to previous subsection, we can expand the potential $V[\phi]$ about the $\phi_{\rm{k}}$. Up to the second order of $\zeta$, one obtains the following expression for the potential:
\begin{equation}\label{poten.}
\begin{array}{cc}
	&	V[\phi]=V[\phi_{\rm{\rm{k}}}]+\int d x\frac{1}{2}\zeta(x) \bigg[-(1+c \phi_{k}^2)\partial^2_{x}- 3c (\partial_x\phi_{k})^2 - 3c (\partial_{x}\phi_{k}^2)\partial_{x}\\&\hspace{3cm}-4 c \phi_{k}(\partial_x^2 \phi_{k})-1+3\phi_{k}^2 \bigg]\zeta(x).
		\end{array}
\end{equation} 
Therefore, the stability equation, read from Eq. \eqref{poten.}, is
\begin{equation}\label{fff}
	\bigg[-(1+c \phi_{k}^2)\partial^2_{x}- 3c (\partial_x\phi_{k})^2 - 3c (\partial_{x}\phi_{k}^2)\partial_{x}-4 c \phi_{k}(\partial_x^2 \phi_{k})-1+3\phi_{k}^2 \bigg]\zeta_n(x)=\omega_n^2\zeta_n(x)
\end{equation}
The above equation can be solved via perturbation on LV parameter $c$. Hence, we expand our states $\zeta_n(x)$ and normal modes $\omega_n$ as
\begin{equation}
	\zeta_n=\zeta_n^{(0)}+c\ {\zeta_n^{(1)}}+c^2{\zeta_n^{(2)}}+\dots,
\end{equation}
\begin{equation}
	\omega_n=\omega_n^{(0)}+c\ {\omega^{(1)}}+c^2{\omega_n^{(2)}}+\dots.
\end{equation}
In the above equations $\zeta_n^{(0)}$ and $
\omega_n^{(0)}$ are the leading order terms given by Eqs. \eqref{BSs} and \eqref{CS}.
Accordingly, the order $c$ of Eq. \eqref{fff} becomes
\begin{eqnarray}\label{steqlv}
	\left\{\partial_x ^2+1+\left(\omega_n^{(0)}\right)^2 -3 \left[\phi _{\rm{k}}^{(0)}(x)\right]^2\right\}\zeta_n ^{(1)}(x)+F_n(x)=0,
\end{eqnarray}
where
\begin{eqnarray}
	&&\nonumber F_n(x)=\bigg[\left(\phi _{\rm{k}}^{(0)}\right)^2\partial_x ^2+3  \partial_x \left(\phi _{\rm{k}}^{(0)}\right)^2\partial_x +2  \omega_n^{(0)} \omega_n^{(1)}\\&&\hspace{1.7cm}-6  \phi _{\rm{k}}^{(0)} \phi _{\rm{k}}^{(1)}+3 \left(\partial_x \phi _{\rm{k}}^{(0)}\right)^2+4 \phi _{\rm{k}}^{(0)} \partial_x ^2\phi _{\rm{k}}^{(0)}\bigg]\zeta_n^{(0)}(x).
\end{eqnarray}
For the two bound states, the solutions of this equation are
\begin{equation}	
	\left\{\begin{array}{cc}
		\hspace{-2.5cm}\zeta^{(1)}_0(x)=	\frac{1}{8} \text{sech}^4\left(\frac{x}{\sqrt2}\right) \left(2 \sqrt{2} x \sinh \left(\sqrt{2} x\right)-9 \cosh \left(\sqrt{2} x\right)+11\right)&\\
		\hspace{6.3cm}\mbox{with}\quad\omega^{(1)}_0=0,\\\hspace{-4.5cm}
		\zeta^{(1)}_1(x)=\frac{1}{20} \text{sech}\left(\frac{x}{\sqrt2}\right) \bigg\{5 \sqrt{2} x \left(1-2 \text{sech}^2\left(x/\sqrt2\right)\right)&\\ \hspace{1.cm}+\tanh \left(\frac{x}{\sqrt2}\right) \left[-2 \sqrt{2} x+4 \log \left(e^{\sqrt{2} x}+1\right)+50\ \text{sech}^2\left(\frac{x}{\sqrt2}\right)-91/3\right]\bigg\}&\\
		\hspace{7.1cm}\mbox{with}\quad \omega^{(1)}_1=\frac{1}{5 \sqrt{6}}.&\\
	\end{array}\right.
\end{equation}
\begin{figure}[h]
	\begin{center} \includegraphics[width=9cm]{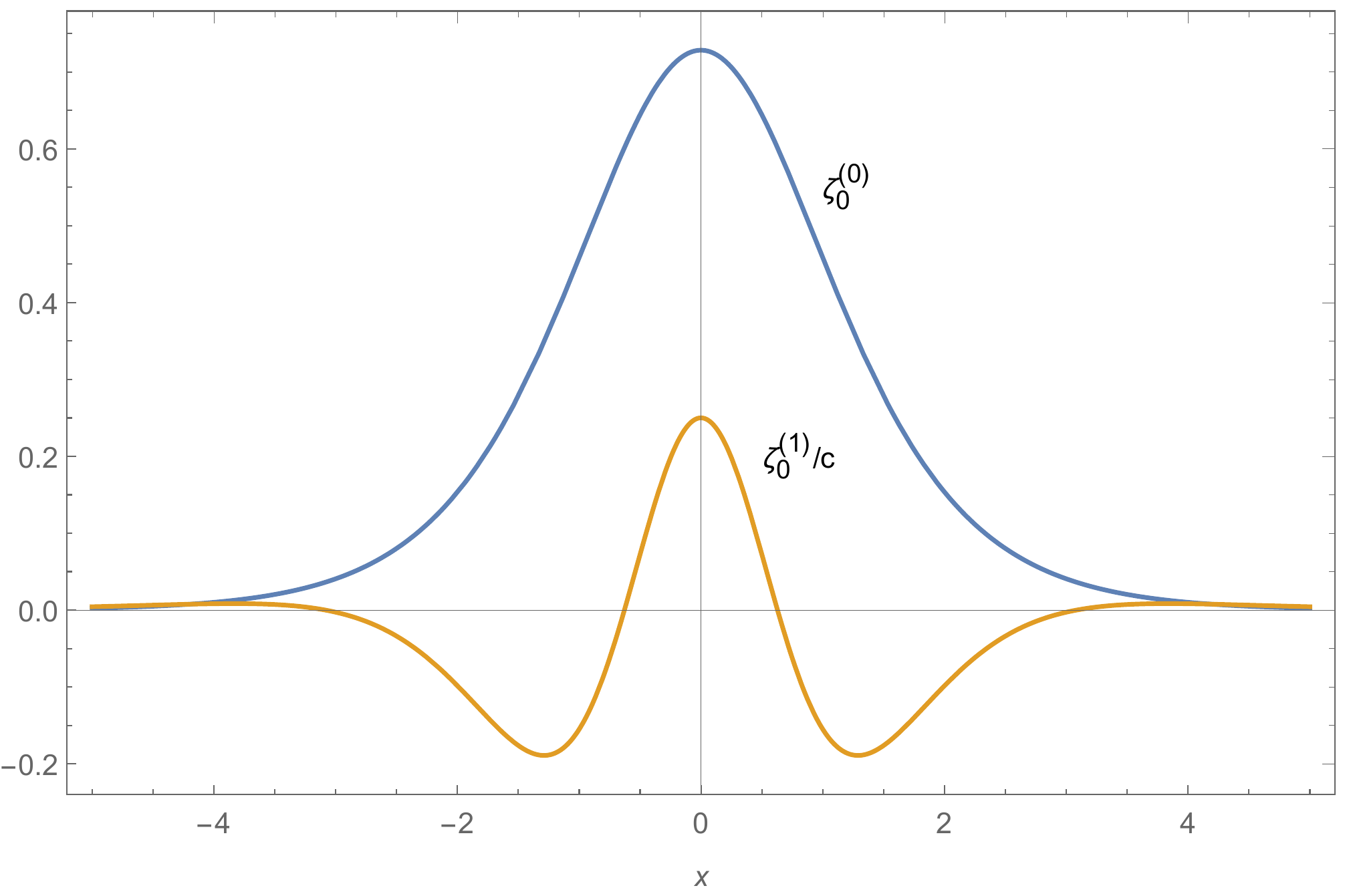}\\\includegraphics[width=9cm]{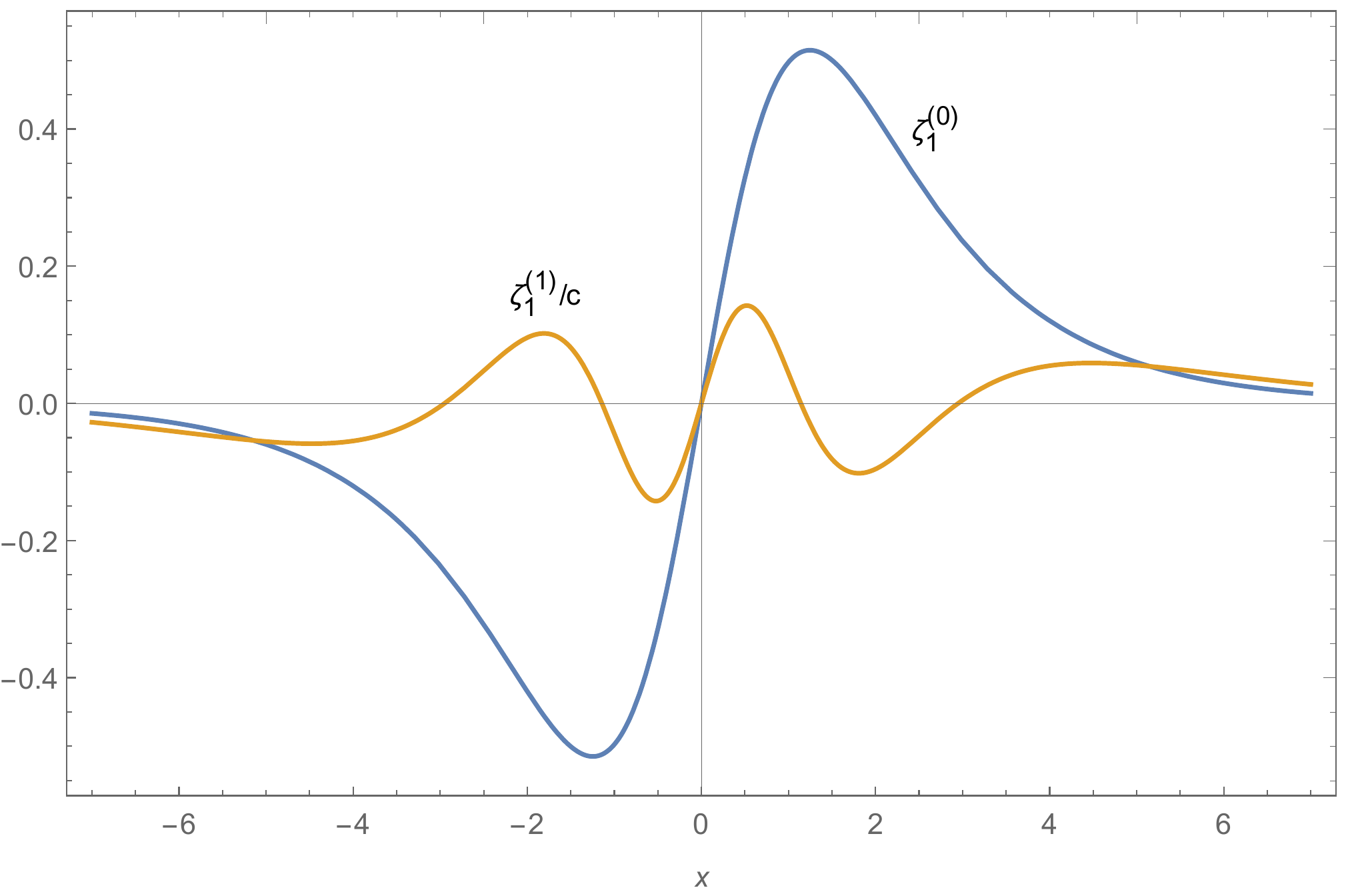} \caption{ {\small The zeroth and first order of two bound states.\label{b0b1}}}
	\end{center}
\end{figure}
Actually, $\omega_0$ does not receive any correction. To derive these solutions and their corresponding modes, we have taken the boundary condition at infinity into account.  For comparison, we have plotted the zeroth and first order bound states in Fig. \ref{b0b1}. For the continuum states, we can obtain corresponding frequencies by writing Eq. \eqref{steqlv} at $x\to\pm \infty$ 
\begin{equation}\label{assf}
	\left(2\partial_x^2 +q^2\right)\zeta^{(1)}(x)+\left(q^2+3 i q-2\right) \left(q^2-2 \sqrt{2} \sqrt{q^2+4} \omega_q^{(1)}\right) e^{i q x/\sqrt2}=0.
\end{equation}
Therefore, we get 
\begin{eqnarray}\label{ffff}
	\nonumber&&\zeta^{(1)}_q(x)=\zeta^{(0)}_q(x)\\&&\qquad
	+Q_2^{i q}\left[\tanh \left(x/\sqrt{2}\right)\right] \int _1^{\tanh \left(x/\sqrt{2}\right)}{{\cal F}_q (t)}P_2^{i q}(t)dt\nonumber\\&&\hspace{2cm}-i P_2^{i q}\left[\tanh \left(x/\sqrt{2}\right)\right] \int _1^{\tanh \left(x/\sqrt{2}\right)}{{\cal F}_q (t)}Q_2^{i q}(t)dt\label{ffff}\\&&
	\quad\mbox{with}\qquad\omega^{(1)}_q=\frac{q^2}{2\sqrt2\sqrt{q^2+4}}.
\end{eqnarray}
where,
\begin{eqnarray}
	\nonumber&&\hspace{-1cm}{{\cal F}_q (t)}=\frac{e^{i q \tanh ^{-1}t}}{(3-i q) \left[P_3^{i q}(t) Q_2^{i q}(t)-P_2^{i q}(t) Q_3^{i q}(t)\right]}\\&& \times[q^4+ \left(-44 q^2-105 i q t+105 t^2-68\right)t^2\nonumber\\&&+6 \left(q^2+3 i q t-3 t^2+1\right) t\tanh ^{-1}t+3 i \left(3 q^2+11\right) q t+4 q^2+3]\nonumber\\
\end{eqnarray}
and $P_n^m(x)$ and $Q_n^m(x)$ are the first and second kinds of the associated Legendre functions, respectively. From Eq. \eqref{assf} we can see that the asymptotic form of the $\zeta^{(1)}_q(x)$ at infinity is $e^{\pm iq x/\sqrt2}$. Therefore we have
\begin{equation}
	\zeta_q(x)\to
	e^{iqx/\sqrt2}\left[2-q^2-3iq+c\right]=\exp[iqx/\sqrt2\pm\delta(q)/2]\quad\mbox{at}\qquad x\to\pm\infty,
\end{equation}
where the phase shift is
\begin{equation}\label{deltalv}
	\delta(q)=-2\tan^{-1}\left(\frac{3q}{2+c-q^2}\right)=-2 \tan ^{-1}\left(\frac{3 q}{2-q^2}\right)+\frac{6  qc}{q^4+5 q^2+4}+{\cal O}(c^2).
\end{equation}
Here, the first term of the right hand side is the leading term of the phase shift (without Lorentz violation) and the second is the correction raised from Lorentz violation. Now, the correction to the energy spectrum becomes
\begin{equation}
	E^{(1)}_{\{n_1,n_{q_i}\}}=
	\frac{\sqrt{2}}{15}+\left(n_1+\frac{1}{2}\right)\frac{1}{5\sqrt{6}}+\sum_{q_i}\left(n_{q_i}+\frac{1}{2}\right)\frac{q_i^2}{2\sqrt2\sqrt{q_i^2+4}}
\end{equation}
where $q_i$ are derived by substituting Eq. \eqref{deltalv} in Eq. \eqref{pbc}.

\section{Summary}\label{summ}

The Lorentz symmetry may
be broken in high energies where the gravity must be quantized.  Therefore, we need to check what causes the
Lorentz violation in low energy theories. In this paper, we have studied the
effects of an LV potential on the energy spectrum predicted for
the $\phi^4$ scalar field theory in two space-time dimensions.
We have considered an additional fundamental interaction  $\frac{-c}{2}\phi^2(u.\partial\phi)^2$  that couples the derivative of a field to a constant vector describing LV. This is the lowest term which leads to nontrivial effects in the theory. The modifications
of wave functions and the corresponding energy spectrum have been investigated
for both trivial vacuum and kink. The trivial vacuum is not affected by the LV potential. However, up to the first order of the LV
coefficient $c$,  we have found the kink and its  excitation modes (all bound and continuum states) are modified. One may expect such modifications because kink is a result of
potential and this new LV term is also in the same order in $\phi$'s.

\vskip20pt\noindent {\large {\bf
		Acknowledgement}}\vskip5pt\noindent
	The authors would like to thank the Research Deputy of the University of Qom for financial support.

\end{document}